\documentclass[12pt]{article}

\ifx\pdfoutput\undefined
\usepackage[dvips,bookmarks]{hyperref}  
\else
\usepackage{hyperref}   
\fi

\usepackage{graphicx}

\hypersetup{colorlinks,bookmarksopen,bookmarksnumbered,citecolor=blue,
        pdfstartview=FitH}
\def\hhref#1{\href{http://arxiv.org/abs/hep-th/#1}{hep-th/#1}} 
\def\mhref#1{\href{mailto:#1}{#1}}              

\def\bop#1{\setbox0=\hbox{$#1M$}\mkern1.5mu
    \vbox{\hrule height0pt depth.04\ht0
    \hbox{\vrule width.04\ht0 height.9\ht0 \kern.9\ht0
    \vrule width.04\ht0}\hrule height.04\ht0}\mkern1.5mu}
\def\bo{{\mathpalette\bop{}}}                        
\def\frac#1#2{{\textstyle{#1\over#2}}}     

\parskip=\medskipamount         
\thicklines             
\begin{document}

\begin{titlepage}
\setcounter{page}{0} 
\begin{flushright}
YITP-SB-06-22\\ June 16, 2006\\
\end{flushright}

\begin{center}
{\centering {\LARGE\bf Random lattice superstrings \par} }

\vskip 1cm {\bf Haidong Feng\footnote{E-mail address:
\mhref{hfeng@ic.sunysb.edu}}, Warren Siegel\footnote{E-mail
address: \mhref{siegel@insti.physics.sunysb.edu}}} \\ \vskip 0.5cm

{\it C.N. Yang Institute for Theoretical Physics,\\
 State University of New York, Stony Brook, 11790-3840 \\}

\end{center}

\begin{abstract}
We propose some new simplifying ingredients for Feynman diagrams that seem necessary for random lattice formulations of superstrings.  In particular, half the fermionic variables appear only in particle loops (similarly to loop momenta), reducing the supersymmetry of the constituents of the Type IIB superstring to N=1, as expected from their interpretation in the 1/N expansion as super Yang-Mills.
\end{abstract}

\end{titlepage}

\section{Introduction}

Strings were introduced originally for hadrons and identified as
bound states of gluons and quarks. Unfortunately, a suitable hadronic
string theory serving that purpose hasn't been constructed. This led
to the reinterpretation of the known strings as fundamental strings
describing gluons and quarks, leptons, gravitons, etc. The
fundamental (super)strings are critically (10)26 dimensional, which
contradicts expectations from QCD, since confinement has critical
dimension 4. There are methods called ``compactification" to
eliminate the extra dimensions, but this leads to the landscape problem of string theory.
(Of course, there are two calculable string theories in 4D: strings
with N=2 worldsheet supersymmetry \cite{M. Ademollo} and twistor
superstrings \cite{V.P. Nair}. But the N=2 superstring has vanishing S-matrices \cite{Oo},
while the twistor string is not a real string and can't describe
physics off shell.)

Replacing the worldsheet by the random lattice is another approach to
string quantization, which directly expresses the string as a bound state of
underlying particles \cite{MR.Douglas}, giving a more precise
correlation between the second-quantization of field theories and
the first-quantization of string theory. The lattice is irregular,
corresponding to the curvature of the worldsheet, and is identified with a Feynman diagram; the functional
integration over worldsheet metrics is the sum over
diagrams \cite{David}.
Presently only the bosonic lattice string is really understood in this formalism. The usual $\frac{1}{2}
(\partial X)^2$ term becomes $\frac{1}{2} (x_i -x_j)^2$, which
produces a propagator $e^{- x^2/2}$ between the two
vertices at $i$ and $j$. The 1/N expansion associates the
faces of the worldsheet polyhedra with the U(N) indices of the
scalar field \cite{Hooft}.  A modification of the random lattice to incorporate standard propagators
describes an asymptotically free theory, ``wrong-sign" $\phi^4$
in 4D \cite{W. Siegel1}.

In the present paper, we consider random lattices for superstring theory.
We will find that general considerations restrict the choice of string formulations and lattice quantizations; these lattices also introduce new kinds of Feynman rules not yet derived from second-quantization of particle theory.
We begin with a brief review of the bosonic string
on the random lattice in the next section.  We then discuss the issues involved in putting fermions on the lattice in a way appropriate to describe superstrings.
In the following section we describe the formalism best suited to these considerations,
the Type IIB superstring in the
infinite ghost pyramid formulation \cite{boo}, which was introduced for covariant
quantization of the superstring. Several examples and general
results for both tree and loop diagrams are
calculated.
The next section describes contributions corresponding to the string measure.
In the conclusions we discuss future directions for
this lattice superstring.
In the appendix we analyze the lattice rules for the Green-Schwarz action of the superstring without gauge fixing.

\section{Bosonic lattice string review}

To find the action of the bosonic string on a lattice, we notice that a
lattice requires a scale, while conformal invariance includes scale
invariance. So the conformal invariance of the worldsheet must be
broken. The simplest scale-variant and coordinate-invariant property
of the worldsheet is its area, so an area term with coefficient (cosmological constant) $\mu$ is added into the
string action. Furthermore, a term containing the string coupling
constant should be included: In string theory, the power of the
coupling constant is counted by the integral of the worldsheet
curvature $R$. So totally, the worldsheet action is
\begin{equation}
  S = \oint {d^2 \sigma\over 2\pi} \sqrt{-g} \left[\frac{1}{\alpha'}
  g^{mn} \frac{1}{2} (\partial_m X \cdot \partial_n X) + \mu + (\ln
  \kappa) \frac{1}{2} R \right]
\end{equation}
On the random lattice, this action can be written as
\begin{equation}
  S_1 = \frac{1}{\alpha'} \sum_{\langle ij \rangle} \frac{1}{2} (x_i - x_j)^2
  + \mu \sum_i 1 + \ln \kappa \left(\sum_i 1 - \sum_{\langle ij \rangle}
  + \sum_J 1\right)
\end{equation}
where $j$ are vertices, $\langle ij \rangle$ the links (edges), and $J$ the
plaquets (faces, planar loops) of the lattice.

The underlying field theory can be found by identifying the
worldsheet lattice with a position-space Feynman diagram. The
vertices and links of the lattice correspond to vertices and
propagators of the Feynman diagram, respectively.  Also, we associate
the 1/N expansion to the faces of the worldsheet polyhedra. The
area term (counting the number of vertices) on the random lattice
gives the coupling constant factor for each vertex in the Feynman
diagram, and the worldsheet curvature term (classifying the
topology) gives the  string coupling 1/N of the topological
expansion. Thus the action of the n-point interaction scalar field
is
\begin{equation}
S_2 = N\ tr \int {d^D x\over (2\pi \alpha')^{D/2}}
\left(\frac{1}{2} \phi e^{-\alpha' \bo /2} \phi - G \frac{1}{n} \phi^n \right)
\end{equation}
with
\begin{equation}
  G = e^{-\mu} , \quad \quad \frac{1}{N} = \kappa
\end{equation}

The Gaussian propagator $e^{-\alpha' \bo /2}$ leads to Gaussian
behavior of fixed-angle scattering, conflicting with the power-law
behavior in hadronic physics. To find ordinary Feynman diagrams,
with $1/p^2$ propagators instead of the Gaussians in the usual
strings, we use the Schwinger parametrization of the propagator
\begin{equation}
 {1\over \frac12 p^2} = \int_0^{\infty} d \tau\ e^{-\tau p^2/2}
\end{equation}
Then the Feynman diagram of a scalar field with nonderivative
interactions can be written as
\begin{equation}\label{Feynmanrule}
  \int d x'_i d p_{ij} d \tau_{ij}\ e^{-\sum_{\langle ij \rangle}
  [\tau_{ij} p_{ij}^2 /2 - i (x_i - x_j) \cdot p_{ij}]}
\end{equation}
where $i, j$ label vertices including those connected to external lines, and $d x'_i$ integrations are over all vertices except them.

We now look for continuum actions that will reproduce the above conventional
Feynman diagrams when the worldsheet is replaced with a random
lattice. The way to construct a continuum action from a random
lattice action is to consider a regular square (``flat" worldsheet)
lattice, and covariantize with respect to the worldsheet metric. In
the (worldsheet) continuum limit of (\ref{Feynmanrule}), $\tau$
must become a symmetric worldsheet tensor.  Since on a regular
square lattice there are two propagators per vertex, the continuum
$\tau$ then has only two components at any point on the worldsheet
and must be a traceless tensor. So by simply setting $\tau_{+-} =0$,
the continuum action is then
\begin{equation}\label{continuum}
  S =  \int{d^2 \sigma\over 2\pi} \sqrt{-g} \left[ i P_{\pm}
  \cdot e^m_{\mp} \partial_m X + \frac{1}{2} \tau_{\pm \pm} P_{\mp}
  \cdot P_{\mp} + \mu + (ln \kappa ) \frac{1}{2} R\right]
\end{equation}
where the ``zweibein" is defined by $g_{mn} = -e_{(m}^+ e_{n)}^-$. It
implies, on the lattice, that the propagators (links) are (worldsheet) lightlike
and the model defined by this theory has only 4-point vertices with
4 lightlike propagators. Thus the scalar field action is
\begin{equation}
S_2 = N\ tr \int{d^D x\over (2\pi \alpha')^{D/2}}
\left(-\frac{1}{4} \phi \bo \phi - G \frac{1}{4} \phi^4 \right)
\end{equation}
This action gives an asymptotically free theory, "wrong-sign"
$\phi^4$ theory in 4D.

\section{Fermions}

There are two related problems with putting fermions on a random lattice: (1) fermion doubling, for worldsheet spinors, which is a problem even for regular lattices (as in lattice QCD) \cite{NN}, and (2) the absence of an unambiguous way to define chiral (left- and right-propagating) fields on a 2D random lattice, since ``left" and ``right" have no clear meanings.

The first problem exists only for worldsheet supersymmetry, but can be resolved for N=2 supersymmetry by ``twisting": redefining (local) Lorentz weights by adding to them (local) U(1) weights, so the worldsheet spinors become tensors.  (Such an approach for regular lattices is called ``K\"ahler-Dirac fermions" \cite{KD}, where the relevant symmetries are global.)

The second problem can be resolved by requiring worldsheet parity invariance.  Then all 2D Levi-Civita tensors can be eliminated, in favor of making any ``pseudotensors" into tensors by absorbing the Levi-Civita tensors into them:  For example, a pseudoscalar becomes a second-rank antisymmetric tensor (2-form).  This restriction on superstrings limits us to Type IIB for spacetime supersymmetric formulations, and N=(2,2) for worldsheet supersymmetric formulations.  (For simplicity, we restrict ourselves to closed strings.  Open strings require independent constituent fields for the boundary and bulk of the worldsheet: e.g., in QCD, quarks live on the boundary and gluons in the bulk.)

Let's first consider free chiral actions.  After twisting (if necessary), each such term will be of the form
$$ A_{++...+}\partial_-B_{-...-} $$
where $A$ has one more ``$+$" than $B$ has $-$'s, accompanied by a similar term related by parity ($+\leftrightarrow -$).  (There might also be overall signs in the parity transformation, but these can be eliminated by field redefinition.)  Each parity doublet then can be easily identified as a totally symmetric, traceless tensor, with an action
$$ A^{mn...p}\partial_m B_{n...p} $$
(The symmetric and antisymmetric combinations of a + with a $-$ would give 2D metric and Levi-Civita tensor, respectively.)

The only exception is the scalar, paired with a pseudoscalar (no indices on $B$).  In that case the sum of left and right $B$'s is the scalar, while the difference becomes a 2-form, with action
$$ A^m (\partial_m B +\partial^n B_{mn}) $$
This is actually the most interesting case, since on a lattice only a scalar is located at a vertex.  For random lattice quantization, this means that only a scalar can be a coordinate of the constituent field.  For example, a vector is identified with a ``link" (propagator), and thus is identified as a momentum, as for the bosonic string, where we have a term $P^m\partial_m X$ in first-order formulation.  But in this case, we also have a 2-form, which is identified with a loop.

The closest analog for the bosonic string is loop momentum:  In a Feynman diagram, if we Fourier transform to introduce momenta in addition to coordinates, varying (integrating over) the coordinates (to eliminate them) imposes momentum conservation at the vertices; solving these conditions replaces the original momenta, defined on each propagator, with loop momenta, defined on each loop.  (This is related to the identity $P-V=L-1$, where the $-1$ represents the remaining unsolved total momentum conservation constraint.)  The corresponding statement in the language of the continuum worldsheet is T-duality \cite{T}: varying $X$ to give $\partial_m P^m=0$ to get $P^m=\partial_n X^{mn}$, where the 2-form $X^{mn}$ is the continuum version of loop momentum.

However, there is some subtlety in the T-duality transformation with respect to zero-modes:  In the random lattice case, this arises when one replaces momenta on external lines with loop momenta, where there is no corresponding loop.  This can be done in a peculiar way by solving also the total momentum conservation constraint by writing the external momenta as differences between ``external-line loop momenta", so that the sum of the differences adds to zero: for $n$ external lines, $p_i=k_i-k_{i+1}$, $k_{n+1}\equiv k_1$.  This introduces a translational invariance in loop momenta that is T-dual to that for the usual spacetime coordinates.  In the continuum worldsheet, these new zero-modes are identified with the winding modes of the string; but for dimensions that are not compactified there are no winding modes, so the interpretation is unclear.

A similar procedure could be applied to the fermion action, T-dualizing the pseudoscalar by using its equation of motion to replace it by a second scalar of which the vector is the gradient.  This might be a viable alternative that we will postpone until a way can be found to deal with the extra zero-modes.

Since free actions can be obtained only after gauge fixing, another alternative would be to avoid gauge fixing and its consequent ambiguities, since that is one of the main advantages of lattice approaches.  For example, for the bosonic string the random lattice replaces integration over the worldsheet metric with summation over Feynman diagrams, in what is apparently a covariant way (or at least as close as one can get with discretization).  For N=2 worldsheet supersymmetry, the U(1) symmetry could be replaced with a compact version, as for regular lattices.  Finally, local supersymmetry does not need to be gauge fixed, because it squares to ``$p^2$", which is not a true constraint:  The usual particle propagator goes as $1/p^2$, not $\delta(p^2)$, so integration over the Lagrange multipliers for supersymmetry should simply produce propagator numerators $\psi\cdot p$, or at worst extra factors of $p^2$ that can be canceled by redefinition of the moduli measure.  Unfortunately, the N=(2,2) string has both a left and a right U(1) symmetry, both gauged by the same worldsheet vector (so that, as usual, Lagrange multipliers are pure gauge, up to moduli), and it is difficult to see how to implement a gauge symmetry whose parameter is a 2-form, since it would transform a vertex variable into a loop variable (unless some type of T-duality transformation were implemented).
Also, there is a combination of N=(2,2) supersymmetries that becomes a scalar after twisting and is nilpotent, thus leading to vanishing amplitudes without gauge fixing.
Such problems are not encountered for the Type IIB superstring; we will consider the Green-Schwarz formulation, without gauge fixing for $\kappa$-symmetry, in the Appendix.

There is a more important reason why gauge fixing is needed:  If we consider the path integral for any amplitude calculation in any continuum-worldsheet formulation of any string theory, at all but a finite number of points on the worldsheet (related to the number of string loops and external lines) only the free terms in the action are used.  (Non-free terms are vertex insertions for external lines, picture-changing operators, contributions from supermoduli, worldsheet instantons, etc.)  If we then discretize this same calculation, we see that there are in general no vertex factors nor propagator numerators in the corresponding Feynman diagram for an arbitrarily large number of vertices and propagators, with the exception of the relatively small number of vertices associated with external lines or string-loop insertions.  This suggests that, at least for formulations of string theories that can be quantized on a random lattice, all such factors should be associated with either: (1) vertex operators for external lines, or (2) string-loop variables coming from particle-loop variables not associated with ``faces" of the worldsheet ``polyhedron" after applying the 1/N expansion (as seen from the identity $L=F+2H-1$, where $H$ is the number of ``handles", or string loops).  (However, the action for even the bosonic string with ordinary propagators for its constituents is not quadratic; here we consider conventional superstrings, from which we assume it is possible to generalize to QCD-like strings with minor modifications.)  Thus, the Feynman rules for the constituent particles should involve only a free, quadratic, first-quantized action, and no vertex factors; only vertex operators for external lines should be nontrivial.

Of course, if the conformal gauge is not fixed, coupling to the worldsheet metric makes the action non-quadratic, but the metric is encoded into the geometry of the random lattice.  In particular, conformal weights are seen in the string action only in terms with the Lorentz connection, which is the derivative of the metric.  (For bosonized fields, the conformal weight appears only in $R\phi$ terms.)  In conformal gauges such contributions are confined to special singular points, such as ``interaction points" or places where external lines are attached.  We can thus treat such contributions in the same way as those considered above.  As a result, we treat all fermions as sets of (worldsheet) vector + scalar + 2-form.  This is natural in that it allows the maximum number of coordinates for the constituent fields, which is half the number expected from the continuum (unless we were to try converting the 2-forms to scalars by T-duality, as discussed above).  Thus, the constituents of the Type IIB superstring have the superspace coordinates of only N=1 supersymmetry.  This is expected from the 1/N expansion, since a U(N) gauge group is associated with super Yang-Mills, which does not allow N=2.  (It also agrees with the AdS/CFT correspondence \cite{AdS}.)

\section{Type IIB superstring}\label{subsection1}

The action for the superstring in the infinite ghost pyramid
formalism, which is a good starting point for covariant
quantization, can be written as
\begin{equation}
S = \int {d^2 \sigma\over 2\pi} \left[\frac{1}{2} (\partial X \cdot
\overline{\partial} X) + \Pi_{1,A} \overline{\partial} \Theta^{1,A}
+ \Pi_{2,A} \partial \Theta^{2,A}\right]
\end{equation}
where we use a general spinor $\Theta^{A}$ ($\Pi_{A}$) to indicate the
usual fermionic coordinates (their conjugate momenta) of superspace,
or the infinite pyramid of ghosts (their conjugates) in the covariant
quantization of the superstring.

\begin{figure}[ht]
\begin{center}
\includegraphics[scale=0.9]{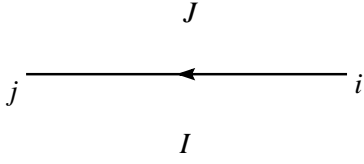}
\end{center}
\caption{\label{fig1} Definition of $\pi_{IJ} \equiv \pi_{ij}$ with
$i,j$ indicating adjacent vertices and $I,J$ adjacent loops on the
lattice.}
\end{figure}

To put this action on the lattice, we have to introduce a worldsheet
scalar $\Theta$ and pseudoscalar $\widetilde{\Theta}$
\begin{eqnarray}\label{inter vertex}
\Theta^{1,A} = {1\over\sqrt 2} (\Theta^A + \widetilde{\Theta}^A) \\
\Theta^{2,A} = {1\over\sqrt 2} (\Theta^A - \widetilde{\Theta}^A)
\end{eqnarray}
Replacing $\Theta^1$ and $\Theta^2$ by $\Theta$ and
$\widetilde{\Theta}$,
\begin{equation}
S = \int {d^2 \sigma\over 2\pi} \left[\frac{1}{2} (\partial^m X
\cdot
\partial_m X) + \Pi^m_A \partial_m \Theta^A +
\Pi^m_A \partial^n  \epsilon_{mn} \widetilde{\Theta}^A\right]
\end{equation}
Thus we can replace the worldsheet with a random lattice,
\begin{eqnarray}\label{4-point amp}
S = \sum_{\langle ij\rangle} \left[\frac{1}{2} (x_i
-x_j)\cdot (x_i -x_j) + \pi_{ij,A} (\theta_i^A - \theta_j^A)\right]
+ \sum_{\langle IJ\rangle}\pi_{IJ,A} (\widetilde{\theta}_I^A -
\widetilde{\theta}_J^A)
\end{eqnarray}
where $i,j,\cdots$ indicate the vertices, $I,J, \cdots$ the (planar)
loops, $\langle ij\rangle$ the links between adjacent vertices, and $\langle IJ\rangle$ the perpendicular (dual) links between adjacent loops:  As shown in Fig.~\ref{fig1}, $\pi_{IJ} \equiv
\pi_{ij}$. Then $\theta_i$ is located at the vertex $i$ and
$\widetilde{\theta}_I$ at the loop $I$.

The Feynman rules derived in this way associate $\tilde\theta$ with only planar loops (faces of the polyhedron), as defined by the 1/N expansion.  However, for field theories of N$\times$N matrices, such as (super) Yang-Mills, the (global) group theory is in no way associated with the (super)coordinates.  Thus, $\tilde\theta$ should appear in the same way for any loop of a Feynman diagram, whether it be planar or nonplanar.  This does not contradict our original assumptions, since the free action did not determine string-loop factors.  The random lattice approach thus predicts $\tilde\theta$'s as string-loop factors.

For tree diagrams $\widetilde{\theta}$ will not be involved. So for the n-point tree, the fermionic part of the path integral is
\begin{eqnarray}\label{npointtree}
{\cal A} = \int \left( \prod_{\mu \nu} d \pi_{A,\mu \nu} \right)\left( \prod_{a} d
\theta_a^A \right)\left[ \prod_{i} d \theta_i^A \phi(x_i, \theta_i)\right]
e^{\pi_{A,\mu \nu} (\theta_\mu^A-\theta_\nu^A)}
\end{eqnarray}
where $\mu,\nu$ are general vertices on lattice, $a$, $i$ internal
and external vertices respectively.

\begin{figure}[ht]
\begin{center}
\includegraphics[scale=0.9]{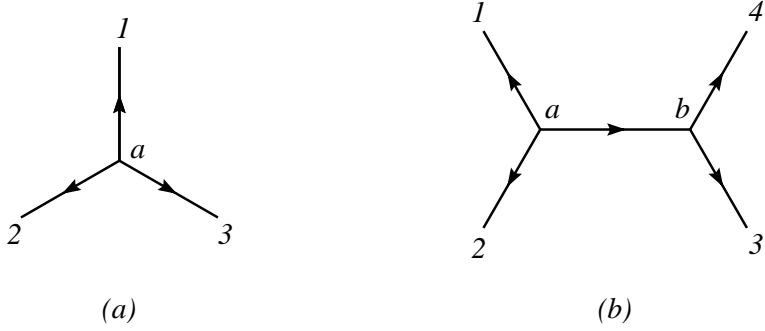}
\end{center}
\caption{\label{fig2} Trees on the lattice. \textit{(a)} 3-point. \textit{(b)} 4-point.}
\end{figure}

The simplest example is the 3-point diagram shown in Fig.~\ref{fig2}a, after
integrating out $\pi_{a1}$, $\pi_{a2}$, and $\pi_{a3}$:
\begin{eqnarray}
{\cal A} & = & \int d \theta_a^A \left[ \prod_{i=1,2,3} d \theta_i^A
\phi(x_i, \theta_i)\right] \left[\prod_A (\theta_a^A-\theta_1^A)\right]
\left[\prod_A (\theta_a^A-\theta_2^A)\right] \left[\prod_A (\theta_a^A-\theta_3^A)\right] \nonumber \\
& = & \int \left[\prod_A (\theta_1^A-\theta_2^A)\right] \left[\prod_A
(\theta_1^A-\theta_3^A)\right] \left[
\prod_{i=1,2,3} d \theta_i^A \phi(x_i, \theta_i)\right] \nonumber \\
& = & \int d \theta^{A} \phi(x_1, \theta) \phi(x_2, \theta)
\phi(x_3, \theta)
\end{eqnarray}
where $\prod_A$ is the antisymmetric product of all components.

Another simple example is the 4-point diagram as shown in Fig.~\ref{fig2}b.
Integrating out $\pi$'s on the links,
\begin{eqnarray}
{\cal A} & = & \int d \theta_a^A d \theta_b^A \left[\prod_{i=1,2,3,4}
d \theta_i^A \phi(x_i, \theta_i)\right] \left[\prod_A
(\theta_a^A-\theta_b^A)\right] \nonumber \\
& & \times \left[\prod_A
(\theta_a^A-\theta_1^A)\right] \left[\prod_A (\theta_a^A-\theta_2^A)\right]
\left[\prod_A (\theta_b^A-\theta_3^A)\right]
\left[\prod_A (\theta_b^A-\theta_4^A)\right] \nonumber \\
& = & \int d \theta^{A} \phi(x_1, \theta) \phi(x_2, \theta)
\phi(x_3, \theta) \phi(x_4, \theta)
\end{eqnarray}

Generalized to any n-point tree graph, integration of $\pi$'s on
links just gives $\delta$ functions between the two $\theta$'s at either
end of that link. Then (\ref{npointtree}) is
\begin{eqnarray}\label{ntree}
{\cal A} & = & \int \left( \prod_{a} d \theta_a^A \right)\left[ \prod_{i} d
\theta_i^A \phi(x_i, \theta_i)\right] \prod_{\mu,\nu}
(\theta_\mu^A-\theta_\nu^A) \nonumber \\
& = & \int d \theta \prod_{i} \phi(x_i, \theta)
\end{eqnarray}
Here $a,b$ and $i,j$ are the internal and external points
respectively, while $\mu,\nu$ are any internal or external points
which are linked to each other.

\begin{figure}[ht]
\begin{center}
\includegraphics[scale=0.9]{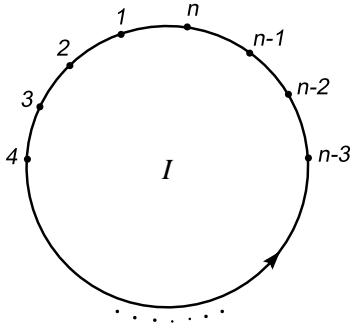}
\end{center}
\caption{\label{fig3} one-loop n-point function on the lattice.}
\end{figure}

For loops, the path integral of the pseudoscalar
$\widetilde{\theta}$ must be done. For a one-loop n-point function as shown in
Fig.~\ref{fig3}, with the spinor index $A$ omitted,
\begin{eqnarray}\label{oneloop}
  {\cal A} & = & \int d \widetilde{\theta} \prod_{i} \left[ d \pi_{i, i+1} d
\theta_i \phi(x_i, \theta_i) e^{\pi_{i,i+1}
(\theta_i-\theta_{i+1})} e^{\pi_{i,i+1}
\widetilde{\theta}} \right] \nonumber \\
& = & \int d \widetilde{\theta} \prod_{n} \left[ d \theta_i \phi(x_i,
\theta_i) (\theta_i - \theta_{i+1} + \widetilde{\theta}) \phi(x_i,
\theta_i)\right]
\end{eqnarray}
The integration of $\widetilde{\theta}$ just replaces
$\widetilde{\theta}$ by $-(\theta_n - \theta_1)$ everywhere. So
(\ref{oneloop}) is
\begin{eqnarray}\label{oneloop1}
 {\cal A} & = & \int \prod_{i=1}^{n-1}
\left\{ d \theta_i \phi(x_i, \theta_i) \left[(\theta_i - \theta_{i+1})
- (\theta_n - \theta_1) \right] \right\} \propto \int d \theta
\prod_{i=1}^{n} \phi(x_{i}, \theta)
\end{eqnarray}

For the general case of n loops, we find a result similar to
(\ref{oneloop1}) with all fermion variables contracted to one. This
is not a surprising result if we notice the relation
\begin{equation}
P-V=L-1 \quad or \quad P-L=V-1
\end{equation}
where $P,V,L$ are the numbers of propagators, vertices, and loops.
Integration over any $\pi_{ij}$ on a link will give a $\delta$-function of
$\theta_i - \theta_j + \widetilde{\theta}_I -\widetilde{\theta}_J$.
For a diagram with $L$ loops, integration over
$\widetilde{\theta}$'s will eliminate $L$ $\delta$-functions. Thus
we are left with $P-L=V-1$ $\delta$-functions and $V$ $\theta$'s to be
integrated, which gives (\ref{oneloop1}).

This result has an obvious interpretation in terms of the zero-modes of the Feynman diagram:  Remember for the worldsheet that any scalar (such as each component of $\theta$ or $\tilde\theta$) has 1 zero-mode, a constant, for any string diagram, while any (2-)vector (such as $\pi^m$) has 2H zero-modes (for H string loops), corresponding to a vector directed around any loop in 2 ways (e.g., consider the torus).  (The vector $P^m$ does not have true zero-modes because of the $P^2$ term, which results in the usual loop momentum integrals.  By ``zero-modes" we here mean modes in the ``de Rham cohomology", closed forms modulo exact forms, i.e., harmonic forms, when those modes are the solutions to the equations of motion.)  A similar result holds for any Feynman diagram:  Instead of using a random lattice approach, we use a first-quantization approach, where any Feynman diagram is treated as a one-dimensional space with a strange topology, and the usual path-integral quantization can be applied \cite{Peng}.  (This is the same sense in which Betti numbers are defined in graph theory.  The variables of the random lattice, i.e., Feynman diagrams, return after solving equations of motion.)  Scalars and vectors of the worldsheet and random lattice have natural generalizations to this 1D space, but 2-forms do not.  Then (each component of) $\theta$ has 1 zero-mode, expressed by the overall $\theta$ integration at the end.  Each $\pi$ has L zero-modes (for L particle loops; e.g., a 1D torus has 1 zero-mode), but they are killed by the L $\tilde\theta$'s.  Since the particle action is 1-dimensional (the worldline), it is trivial except for the zero-modes, which are now also trivial.  (For $X$, the only nontrivial part is the integration over the would-be zero-modes, the loop momenta.)

To relate the Feynman diagrams to the string, we must take into account the 1/N expansion.  Then the particle loops must be divided up into those associated with faces and those not.  Just as the string is derived by summing the perturbation expansion in $Ng^2$ and not that in the string coupling $1/N$ (ignoring subtleties of worldsheet renormalization), the superstring is also derived by integrating over $\tilde\theta$'s associated with faces and not over those associated with string loops.  Since $L=F+2H-1$, that leaves $2H-1$ $\tilde\theta$'s, $2H$ to kill the string's $\pi$ zero-modes, less 1 because we have already killed 1 by integrating out $\tilde\theta$'s zero-modes.  (On the continuum worldsheet it is more natural to think of both $\theta$ and $\tilde\theta$ as coordinates, manifesting N=2 spacetime supersymmetry.)

A similar approach can be applied to the gauge-fixed, twisted N=(2,2) worldsheet supersymmetric formulation of the superstring (including the bosonic ghosts that are partners to the fermions, and the fermionic ghosts for U(1)).  The constituent particles again have half the supersymmetry, describing an N=2 spinning particle, and thus Yang-Mills \cite{Town}.  Presumably this particle is also selfdual, as this string is known to be \cite{Oo}.

\section{Measure considerations}

In the previous section we evaluated the dependence of general Feynman diagrams on external fermions.  This corresponds to the exponentiated worldsheet-Green-function contributions to the path integral.  In any Feynman diagram, in addition to dependence on external momenta and fermions, there are also numerical factors coming from the usual combinatorics.  In the continuum string these appear as functional determinants.  They depend only on the moduli of the worldsheet metric, i.e., the (conformal) geometry.  On the random lattice, the dependence is on the ``geometry" of the graph.

The relation between any such combinatoric factors to string expressions is not obvious, since it requires an understanding of the continuum limit, which may be nonperturbative, or at least the identification of which graph properties will survive in this limit as worldsheet moduli.  However, although the explicit expression for these factors may be tedious, the general form is easy to describe:  In particular, we can simply compare the fermionic contribution to the bosonic one, and see the analog of cancelation of their ``partition functions".

The only equation is the graphical analog of the worldsheet ``curl".  Such an expression arises, e.g., in solving the conservation-of-momentum condition that follows from integrating out $X$ (or $x_i$ on the lattice):
$$ \partial_m P^m = 0 \quad\Rightarrow\quad P^m = \epsilon^{mn}\partial_n \tilde P $$
as in T-duality discussed above, becomes on the lattice
$$ \sum_j p_{ij} = 0 \quad\Rightarrow\quad p_{ij} = \sum_I c_{Iij}k_I $$
for some constants $c_{Iij}$ that depend only on the geometry of the graph, where $k_I$ are the usual loop momenta.  (This relation can be made trivial for planar graphs, as $p_{ij}=k_I-k_J$, but functional determinants are rather trivial for such string-tree graphs anyway.)  Clearly the same expressions will result for the fermions upon integrating out $\Theta$:
$$ \partial_m \Pi^m = 0 \quad\Rightarrow\quad \Pi^m = \epsilon^{mn}\partial_n \tilde\Pi $$
$$ \sum_j \pi_{ij} = 0 \quad\Rightarrow\quad \pi_{ij} = \sum_I c_{Iij}\tilde\pi_I $$
However, we have the same translation for the curl in the $\tilde\Theta$ part of the action:
$$ \int \Pi^m \epsilon_m{}^n \partial_n \tilde\Theta \quad\rightarrow\quad
\sum_{I,\langle ij\rangle} \pi_{ij} c_{Iij} \tilde\theta_I $$
The surviving terms in the action (except those for external variables) are then
$$ \sum_{IJ} (\frac12 k_I M_{IJ} k_J + \tilde\pi_I M_{IJ} \tilde\theta_J ) $$
where
$$ M_{IJ} = \sum_{\langle ij\rangle}c_{Iij}c_{Jij} $$
is essentially the worldsheet d'Alembertian defined on loops.  Final integration of the loop momenta and fermions then yields determinants for this ``kinetic operator":
$$ (det\ M)^{4-D/2} $$
where the 4=16/4 for the fermions comes from the usual $1-2+3-4+...=1/4$ for the ghost pyramid's 16-component spinors.  There would be exact cancelation in $D=10$ except for the fact that the random lattice does not fix worldsheet coordinate invariance, so there is no contribution from corresponding ghosts; that role is played by integration over the worldsheet metric, which for the random lattice means summation over all graphs.

\section{Conclusions}

We have examined the random lattice formulation of the Type IIB superstring in the ghost pyramid formalism.  The constituent particles are functions of the same variables as in the ghost pyramid formulation of the N=1 superparticle (super Yang-Mills).  However, a new type of variable appears in the superparticle Feynman rules, a loop fermion analogous to loop momenta, but not resulting from Fourier transformation of the superparticle coordinates.

The treatment given here should be sufficient for the ``free" part (from the continuum point of view) of random lattice quantization.  However, the construction of vertex insertions for external lines is not obvious, even though our description already includes internal vertices, for several reasons:  (1) To obtain scattering amplitudes for the usual string states, one needs vertex operators for fields that are composite with respect to the constituent fields (color-singlets with respect to the U(N) of the 1/N expansion).  (2) The rules obtained may be for a background-field gauge, as expected from a string approach, which has different external and internal vertices.  (3) The $\tilde\theta$'s that we have derived from the string, but not from particle field theory, may be the result of manipulations of internal vertex factors, obscuring their origin.  (4) The simple form of the Feynman rules we have derived may apply only to the usual superstrings, and QCD-like strings may require more conventional and more complicated (but perhaps not too much) vertex factors.

This approach should be useful not only for string theory, but also for conventional particle field theory.  In particular, it is important to discover how loop fermions such as $\tilde\theta$ can arise from second quantization.  Such rules could significantly simplify loop calculations in maximally supersymmetric field theories, as long expected from the fact that the results of such calculations are much simpler than in theories with less symmetry.

\section*{Acknowledgement}

This work is supported in part by National Science Foundation
Grant No.\ PHY-0354776.

\appendix

\section{Green-Schwarz formalism}\label{subsection2}

As in section \ref{subsection1}, we introduce a scalar $\Theta$ and
pseudoscalar $\widetilde{\Theta}$, located at vertices and loops on
the lattice respectively,
\begin{eqnarray}\label{inter vertex again}
\Theta^{1,A} = {1\over\sqrt 2}( \Theta^A + \widetilde{\Theta}^A ) \\
\Theta^{2,A} = {1\over\sqrt 2}( \Theta^A - \widetilde{\Theta}^A )
\end{eqnarray}
(For alternative attempts at dealing with Levi-Civita tensors on the
worldsheet for Green-Schwarz, see \cite{randomsuper}.) Then the
action for the Type IIB superstring in the Green-Schwarz action
\cite{GS} can be expanded in $\tilde\Theta$ as
\begin{equation}\label{GS action1}
 S = \int {d^2 \sigma\over2\pi}  
 ( {\cal L}_0 + {\cal L}_1 + {\cal L}_2 + {\cal L}_3 + {\cal L}_4 )
\end{equation}
with
\begin{eqnarray}\label{L}
 {\cal L}_0 &=& \frac12 (\partial X -\Theta\gamma\partial\Theta)^2
 \nonumber\\
 {\cal L}_1 &=& \epsilon^{mn} (\partial_m X -\frac12 \Theta \gamma \partial_m \Theta) 
\cdot (\Theta \gamma\partial_n \widetilde{\Theta}
+ \widetilde{\Theta} \gamma \partial_n \Theta) 
 \nonumber\\
 {\cal L}_2 &=& 
- (\partial X - \Theta \gamma \partial\Theta) \cdot
(\widetilde{\Theta} \gamma \partial \widetilde\Theta) 
 \nonumber\\
{\cal L}_3 &=& -\frac12\epsilon^{mn}(\widetilde{\Theta} \gamma
\partial_m \widetilde{\Theta}) \cdot
(\widetilde{\Theta}\gamma\partial_n \Theta + \Theta \gamma \partial_n
\widetilde{\Theta})
 \nonumber\\
{\cal L}_4 &=& \frac12(\widetilde{\Theta} \gamma\partial\widetilde{\Theta})^2
\end{eqnarray}

For tree graphs, terms with $\widetilde{\Theta}$ won't be involved
because $\widetilde{\Theta}$'s are the variables located in loops.
Replacing the worldsheet with the random lattice,
\begin{equation}
  S_0 =
  \frac{1}{2} \sum_{\langle ij \rangle} [
  (x^{\mu}_i - x^{\mu}_j) + \theta_i \gamma^{\mu} \theta_j]^2
\end{equation}
In first-order formalism, it is
\begin{eqnarray}
  S_0 &=& \sum_{\langle ij \rangle} [- \frac{1}{2} p_{ij}^2 +
p_{ij}\cdot (x_i-x_j + \theta_i \gamma \theta_j )]
\end{eqnarray}
We now use the identity for any function $f$
\begin{equation}
 f(x-x' + \theta \gamma \theta') = d^{D'} f(x-x') \delta^{D'}
(\theta - \theta')
\end{equation}
where $d$ is the covariant spinor derivative and $d^{D'}$ the
antisymmetric product of all its components. Then the propagator is
(with appropriate 2D Wick rotation)
\begin{eqnarray}\label{tree}
  \Delta_{ij} = e^{-S_0} = e^{- p_{ij}^2/2} d_i^{D'}
\left[e^{ip_{ij}\cdot (x_i-x_j)} \delta^{D'} (\theta_i
- \theta_j) \right]
\end{eqnarray}
Except for the Gaussian factor $e^{- p^2/2}$ instead of
$1/\frac12 p^2$, this is the same expression that follows from path-integral
quantization \cite{path} of the Casalbuoni-Brink-Schwarz
superparticle action \cite{CBS} without gauge fixing
$\kappa$-symmetry.  In the case D=3, this result agrees with usual
supergraph rules for the N=1 scalar multiplet. Thus, except for this
replacement, the Feynman rules for this multiplet agree with the 3D
N=1 superstring theory obtained by dropping the WZ term (and thus
avoiding the need for a $\Theta$-$\tilde\Theta$ split). Because of
the degeneracy of 3D N=1 superspace (where $\theta$ has only 2
components, so the usual statement that only 1/4 of all $\theta$'s
are physical can't apply), in that case the WZ term of the usual GS
action is unnecessary for continuum quantization \cite{3Dstring}.

\begin{figure}[ht]
\begin{center}
\includegraphics[scale=0.9]{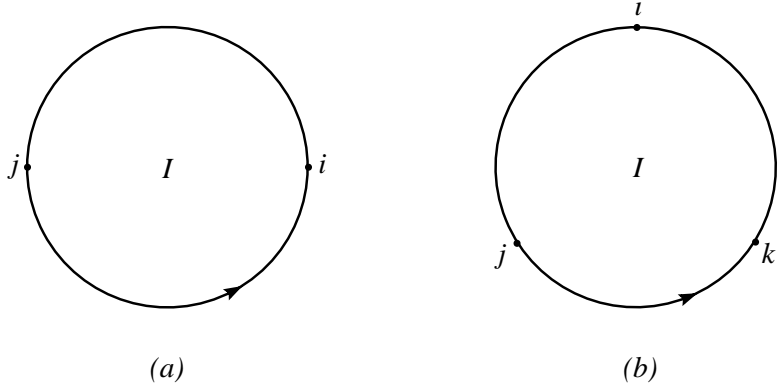}
\end{center}
\caption{\label{fig4} One loop on the lattice. \textit{(a)} 2-point.
\textit{(b)} 3-point.}
\end{figure}

For one-loop diagrams, only terms with one $\widetilde{\Theta}$ are
involved,
\begin{equation}
{\cal L}_1 = - 2\epsilon^{mn} [\partial_m X+ (\partial_m\Theta) \gamma \Theta] 
 \cdot (\partial_n \Theta)\gamma \widetilde{\Theta}
\end{equation}
where we have simplified (\ref{L}) by partial integration and using the
identity
\begin{equation}
  \eta_{\mu \nu} \gamma^{\mu}_{A(B} \gamma^{\nu}_{CD)} =0
\end{equation}
On the random lattice, $\partial X \rightarrow x_i -x_j$ and
$\partial \Theta \rightarrow \theta_i - \theta_j$. The one-loop
two-point diagram, as shown in Fig.~\ref{fig4}a, vanishes. Now
consider the one-loop 3-point diagram, as shown in Fig.~\ref{fig4}b.
Defining
$$\hat p^{\mu}_{ij} \equiv x^{\mu}_i- x^{\mu}_j + \theta_i \gamma^{\mu} \theta_j$$
($p$ on shell) we have
\begin{equation}
S_1 = -2 [ (\hat p_{ij} + \hat p_{ik})(\theta_j - \theta_k) + (\hat p_{jk} + \hat p_{ji})
(\theta_k - \theta_i) 
 + (\hat p_{ki} + \hat p_{kj}) (\theta_i - \theta_j) ] \cdot \gamma \widetilde{\theta}
\end{equation}
Integration over $\widetilde{\theta}$ gives the amplitude
\begin{eqnarray}
{\cal A} = \frac{1}{3} \int d \theta_i d \theta_j d \theta_k
 d x_i d x_j d x_k d p_{ij} d p_{jk} d p_{ki} \Delta_{ij} \Delta_{jk} \Delta_{ki}
\phi(x_i,\theta_i) \phi(x_j,\theta_j) \phi(x_k,\theta_k) \Phi
\end{eqnarray}
where $\Delta_{ij}$ is the propagator in (\ref{tree}) and
\begin{equation}
 \Phi = 2^{D'} \prod_A\gamma_{AB}\cdot [ (\hat p_{ij} + \hat p_{ik})
(\theta_j - \theta_k)^B + (\hat p_{jk} + \hat p_{ji})
(\theta_k - \theta_i)^B + (\hat p_{ki} + \hat p_{kj})
(\theta_i - \theta_j)^B ] 
\end{equation}
where $\prod_A$ is the antisymmetric product of all
components and $D'$ is the range of the
spinor index $A$.  (For 3D, $D'=2$.)

As in the usual
supergraph calculations, we then can do partial integration of all the
$d_i$'s in each $\Delta_{ij}$ until we are left with
\begin{eqnarray}
  e^{-p_{ij}^2/2} e^{ip_{ij} \cdot (x_i-x_j)}
  \delta^{D'} (\theta_i - \theta_j) = \Delta_{b,ij} \delta^{D'} (\theta_i - \theta_j)
\end{eqnarray}
where $\Delta_{b,ij}$ is the bosonic propagator. Let's consider this
procedure for the simple 3D case, with two-component spinor index
$A$.  (We use the conventions of \cite{superspace}, supplemented with the antihermitian $\gamma$ matrices $\{ d_A, d_B \} = -2\gamma_{AB}\cdot\partial$ for comparison with higher dimensions.) After the
partial integration of $d_i$'s off $\Delta_{ij}$,
\begin{eqnarray}
{\cal A} &=& \frac{1}{3} \int d \theta_i d \theta_j d \theta_k
 d x_i d x_j d x_k dp_{ij} d p_{jk} d p_{ki} {\cal B}
\end{eqnarray}
with (using $d^2=\frac12 d^A d_A$)
\begin{eqnarray}\label{3D1loop}
{\cal B} &=&
\Delta_{b,ij} \delta^{2} (\theta_i - \theta_j) [ \Delta_{jk}
\Delta_{ki} (d_i^2 \phi_i) \phi_j \phi_k \Phi
+ \Delta_{jk}(d_i^2\Delta_{ki}) \phi_i \phi_j \phi_k \Phi
\nonumber \\
&& \quad \quad \quad \quad \quad \quad
+ \Delta_{jk}(d_i^A \Delta_{ki}) (d_{i,A} \phi_i) \phi_j \phi_k \Phi
+ \Delta_{jk}\Delta_{ki} \phi_i \phi_j \phi_k (d_i^2 \Phi)
\nonumber\\
&& \quad \quad \quad \quad \quad \quad +
\Delta_{jk}(d_i^A \Delta_{ki}) \phi_i \phi_j \phi_k (d_{i,A} \Phi)
+ \Delta_{jk}\Delta_{ki} (d_i^A \phi_i) \phi_j \phi_k (d_{i,A} \Phi) ]
\end{eqnarray}
For the terms where $d_i$'s don't act on $\Phi$, because $\theta^i =
\theta^j$,
\begin{eqnarray}
 \Phi = 6^2 \prod_A [ \gamma_{AB}\cdot (x_i-x_j)
(\theta_j - \theta_k)^B ] 
= 6^2 \
det[\gamma\cdot (x_i-x_j)] \delta^2 (\theta_j -\theta_k)
\end{eqnarray}
Combining the delta function $\delta^2 (\theta_j - \theta_k)$
here with $\Delta_{jk}$, we notice
$$\delta^2 (\theta_j - \theta_k)\Delta_{jk} = \Delta_{b,jk} \delta^2 (\theta_j - \theta_k)$$ 
The first term in (\ref{3D1loop}) is then, by noticing
$det[ \gamma\cdot (x_i-x_j)]= (x_i-x_j)^2$,
\begin{eqnarray}
 && 6^2 \Delta_{b,ij} \delta^{2} (\theta_i - \theta_j) \Delta_{b,jk}
 \delta^{2} (\theta_j - \theta_k) \Delta_{ki} (d_i^2 \phi_i)
 \phi_j \phi_k  det[(x_i-x_j)_{\mu} \gamma^{\mu}_{BA}] \nonumber \\
 &=& 6^2 \Delta_{b,ij} \Delta_{b,jk} \Delta_{b,ki} \delta^{2} (\theta_i - \theta_j)
 \delta^{2} (\theta_j - \theta_k) (d_i^2 \phi_i)
 \phi_j \phi_k (x_i-x_j)^2
\end{eqnarray}
while the second and third terms just vanish by noticing
$$\delta^2
(\theta_k - \theta_i) d_{A,i} \Delta_{ki}= \delta^2 (\theta_k -
\theta_i) d_i^2 \Delta_{ki}= 0$$

For the last three terms, we write
$$\Delta_{jk} = e^{- p_{kj}^2/2} d_k^2 [e^{i
p_{kj} (x_k-x_j)} \delta^{2} (\theta_k - \theta_j) ]$$
and integrate this $d_k^{2}$ by parts again. For the fourth
term, $d_k$'s can't act on $\Delta_{ki}$. Thus it is
\begin{eqnarray}\label{term4}
 && \Delta_{b,ij} \Delta_{b,jk} \Delta_{b,ki} \delta^{2} (\theta_i -
\theta_j) \delta^{2} (\theta_j - \theta_k) \phi_i \phi_j d_k^2 (
\phi_k d_i^2 \Phi) \nonumber \\&=& \Delta_{b,ij} \Delta_{b,jk}
\Delta_{b,ki} \delta^{2} (\theta_i - \theta_j) \delta^{2} (\theta_j
- \theta_k) \phi_i \phi_j (d_k^2 \phi_k) (d_i^2 \Phi)
\end{eqnarray}
Here we reach the second step by noticing that only $\delta^{2} (\theta_i
- \theta_j) \delta^{2} (\theta_j - \theta_k) (d_i^2 \Phi)$ survives
because
$\hat p^{\mu}_{ij} = d_i^2 [ (x^{\mu}_i- x^{\mu}_j) \delta^2
(\theta_i - \theta_j) ] $ and $d_i d_i^2 = \partial_{i, \mu}
\gamma^{\mu}_{AB} d^i_B, d_i^2 d_i^2 = (\partial_{i})^2$.
For the
same reason, the fifth term is
\begin{eqnarray}\label{term5}
\Delta_{b,ij} \delta^{2} (\theta_i -
\theta_j)  \Delta_{b,jk} \delta^{2} (\theta_j - \theta_k) (d_k^A
d_i^B \Delta_{ki}) \phi_i \phi_j \phi_k (d_{k,A} d_{i,B} \Phi)
\end{eqnarray}
and the sixth term is
\begin{eqnarray}\label{term6}
\Delta_{b,ij} \delta^{2} (\theta_i -
\theta_j) \Delta_{b,jk} \delta^{2} (\theta_j - \theta_k) \Delta_{ki}
(d_i^A \phi_i) \phi_j (d_k^B \phi_k) (d_{k,B} d_{i,A} \Phi)
\end{eqnarray}
Then it is easy to find
\begin{eqnarray}
  d_i^2 \Phi|_{\theta_i = \theta_j= \theta_k} = 6^2 \ det[\gamma\cdot (x_j -
  x_k) ] = 6^2 (x_j - x_k)^2
\end{eqnarray}
and
\begin{eqnarray}
  (d^k_D d^i_B \Phi)|_{\theta_i = \theta_j= \theta_k} =
  \frac{1}{2} 6^2 C_{BD} (x_i -x_j)_{\mu} (x_j -x_k)^{\mu}
\end{eqnarray}
So (\ref{term4}) is
\begin{eqnarray}
 6^2 \Delta_{b,ij} \Delta_{b,jk} \Delta_{b,ki} \delta^{2} (\theta_i - \theta_j) \delta^{2} (\theta_j
- \theta_k) \phi_i \phi_j (d_k^2  \phi_k) (x_j - x_k)^2
\end{eqnarray}
For (\ref{term5}), using the familiar relation $d_C d_A d^2 =
\partial_{\mu} \gamma^{\mu}_{CA} d^2$ +  terms with no $d$ (which may be
dropped), it vanishes because of the factor
\begin{eqnarray}
  C^{AB} C^{CD}  C_{BD} \gamma^{\mu}_{CA} = 0
\end{eqnarray}
It is also easy to evaluate (\ref{term6}), which is
\begin{eqnarray}
\frac{1}{2} 6^2 \Delta_{b,ij} \Delta_{b,jk} \Delta_{b,ki} \delta^{2}
(\theta_i - \theta_j) \delta^{2} (\theta_j - \theta_k) (d_i^A
\phi_i) \phi_j (d_{k,A} \phi_k) (x_i -x_j)\cdot (x_j - x_k)
\end{eqnarray}

Finally,
\begin{eqnarray}
{\cal A} &=& \frac{1}{3} 6^2 \int d x^i d x^j d x^k d p_{ij} d
p_{jk} d p_{ki} d \theta \Delta_{b,ij} \Delta_{b,jk} \Delta_{b,ki}
[(d^2 \phi_i)
 \phi_j \phi_k (x_i-x_j)^2 \nonumber \\ && \quad + \phi_i \phi_j (d^2  \phi_k) (x_j -
x_k)^2 + \frac{1}{2}(d^A \phi_i) \phi_j (d_A \phi_k) (x_i
-x_j)_{\mu} (x_j - x_k)^{\mu} ]
\end{eqnarray}

Generalizing to 10D with 16-component spinor indices, the
calculation is pretty much similar. For a vector $M_{AB} = M_{\mu}
\gamma^{\mu}_{AB}$, using the determinant definition
\begin{equation}
  det M_A^B = \frac{1}{16!} \epsilon^{A_1 \cdots A_{16}} \epsilon_{B_1 \cdots B_{16}}
  M_{A_1}^{B_1} \cdots M_{A_{16}}^{B_{16}} = (M^2)^8
\end{equation}
the one-loop 3-point amplitude is
\begin{eqnarray}
{\cal A} &=& \frac{1}{3} 6^{10} \int d x^i d x^j d x^k d p_{ij} d
p_{jk} d p_{ki} d \theta \Delta_{b,ij} \Delta_{b,jk} \Delta_{b,ki}
\nonumber
\\ && \quad \quad \times \{ [(d^{16} \phi_i)
 \phi_j \phi_k [(x_i-x_j)^2]^8 + \phi_i \phi_j (d^{16}  \phi_k) [(x_j -
x_k)^2]^8 \nonumber \\ && \quad \quad \quad \quad + \sum_{m=1}^{15}
\epsilon^{A_1 \cdots A_{16}} [(d^m_{A_1 \cdots A_m} \phi_i) \phi_j
(d^{16-m}_{A_{m+1} \cdots A_{16}} \phi_k) J_m] \}
\end{eqnarray}
where $J_m$ is the coefficient of $\lambda^m$ in $\det [
(x_k-x_j)_{\mu} \gamma^{\mu} - \lambda (x_j-x_i)_{\mu}
\gamma^{\mu}]$.


\begin{thebibliography}{99}

\bibitem{M. Ademollo}
M. Ademollo, L. Brink, A. D'Adda, R. D'Auria, E. Napolitano, S.
Sciuto, E. Del Giudice, P. Di Vecchia, S. Ferrara, F. Gliozzi, R.
Musto, R. Pettorino, and J.H. Schwarz, {\it Nucl. Phys. B} {\bf 111}
(1976) 77.

\bibitem{V.P. Nair}
V.P. Nair, {\it Phys. Lett.} {\bf 214B} (1988) 215; \\
E. Witten, \hhref{0312171}, {\it Comm.Math.Phys.} {\bf 252} (2004) 189; \\
R. Roiban, M. Spradlin, and A. Volovich, \hhref{0402016}, {\it JHEP}
{\bf 0404} (2004) 012, \hhref{0403190}, {\it Phys.Rev. D} {\bf 70} (2004) 026009; \\
R. Roiban and A. Volovich, \hhref{0402121}, {\it Phys.Rev.Lett.} {\bf  93} (2004) 131602; \\
N. Berkovits, \hhref{0402045}, {\it Phys.Rev.Lett.} {\bf 93} (2004) 011601.

\bibitem{Oo}
H. Ooguri and C. Vafa, {\it Nucl. Phys.} {\bf B361} (1991) 469;
{\it Mod. Phys. Lett.} {\bf A5} (1990) 1389.

\bibitem{MR.Douglas}
M.R. Douglas and S.H. Shenker, {\it Nucl. Phys.} {\bf B335} (1990) 635; \\
D.J. Gross and A.A. Migdal, {\it Phys. Rev. Lett.} {\bf 64} (1990) 125; \\
E. Br\'{e}zin and V.A. Kazakov, {\it Phys. Lett.} {\bf 263B} (1990) 144.

\bibitem{David}
F. David, {\it Nucl. Phys.} {\bf B257} [FS14] (1985) 543; \\
V.A. Kazakov, I.K. Kostov, and A.A. Migdal, {\it Phys. Lett.} {\bf 157B} (1985) 295;\\
J. Ambj\o rn, B. Durhuus, and J. Fr\"ohlich, {\it Nucl. Phys.} {\bf B257} (1985) 433.

\bibitem{Hooft}
G. 't Hooft, {\it Nucl. Phys.} {\bf B72} (1974) 461.

\bibitem{W. Siegel1}
W. Siegel, \hhref{9601002}, {\it Int. J. Mod. Phys. A} {\bf 13} (1998) 381.

\bibitem{boo}
K. Lee and W. Siegel, \hhref{0506198}, {\it JHEP} {\bf 0508} (2005) 102, \hhref{0603218}.

\bibitem{NN}
H.B. Nielsen and M. Ninomiya, {\it Nucl. Phys.} {\bf B185} (1981) 20, {\bf 195} (1982) 541, 
{\bf 193} (1981) 173, {\it Phys. Lett.} {\bf 105B} (1981) 219.

\bibitem{KD}
D. Ivanenko and L. Landau,  {\it Z. Physik} {\bf 48} (1928) 340;\\
C. Lanczos, {\it Z. Physik} {\bf 57} (1929) 447, 474, 484;\\
A.W. Conway, {\it Proc. Roy. Soc. A} {\bf 162} (1937) 145;\\
E. K\"ahler, {\it Rendiconti di Matematica} (3-4) {\bf 21} (1962) 425.

\bibitem{T}
W. Siegel, {\it Phys. Lett.} {\bf 134B} (1984) 318;\\
T.H. Buscher, {\it Phys. Lett.} {\bf 194B} (1987) 59, {\bf 201B} (1988) 466;\\
W. Siegel, {\it Phys. Lett.} {\bf 252B} (1990) 558.

\bibitem{AdS}
L. Maldacena, \hhref{9711200}, {\it Adv. Theor. Math. Phys.} {\bf 2} (1998) 231;\\
S.S. Gubser, I.R. Klebanov, and A.M. Polyakov,
\hhref{9802109}, {\it Phys. Lett.} {\bf B248} (1998) 105;\\
E. Witten, \hhref{9802150}, {\it Adv. Theor. Math. Phys.} {\bf 2} (1998) 253.

\bibitem{Peng}
P. Dai and W. Siegel, in preparation.

\bibitem{Town}
P.S. Howe, S. Penati, M. Pernici and P.K. Townsend, {\it Phys. Lett.} {\bf 215B} (1988) 555,
{\it Class. Quant. Grav.}  {\bf 6} (1989) 1125.

\bibitem{randomsuper}
A. Mikovi\'c and W. Siegel, {\it Phys. Lett.} {\bf 240B} (1990) 363;\\
W. Siegel, \hhref{9403144},  {\it Phys.Rev. D} {\bf 50} (1994) 2799;\\
S. Oda and T. Yukawa, \hhref{9903216}, {\it Prog. Theor. Phys.} {\bf 102} (1999) 215, 
\href{http://arxiv.org/abs/hep-lat/9912008}{hep-lat/9912008},
{\it Nucl. Phys. Proc. Suppl.} {\bf 83} (2000) 754.

\bibitem{GS}
M.B. Green and J.H. Schwarz, {\it Phys. Lett.} {\bf 136B} (1984) 367, 
{\it Nucl. Phys.} {\bf  B243} (1984) 285.

\bibitem{path}
W. Siegel, {\it Phys. Lett.} {\bf 128B} (1983) 397.

\bibitem{CBS}
R. Casalbuoni, {\it Phys. Lett.} {\bf 62B} (1976) 49;\\
L. Brink and J.H. Schwarz, {\it Phys. Lett.} {\bf 100B} (1981) 310.

\bibitem{3Dstring}
W. Siegel, {\it Nucl. Phys.} {\bf B236} (1984) 311;\\
J. Ambj\o rn and S. Varsted, {\it Phys. Lett.} {\bf 257B} (1991) 305.

\bibitem{superspace}
S.J. Gates, Jr., M.T. Grisaru, M. Ro\v cek, and W. Siegel, \hhref{0108200},
{\it Front. Phys.} {\bf 58} (1983) 1.

\end{thebibliography}
\end{document}